\documentclass[]{article} 

\usepackage{lpsc_style} 
\usepackage{times}
\usepackage{journals}

\name{Raquel B. Salmeron} 
\ownemail{raquel@mso.anu.edu.au}
\DocTitle{Chondrule formation via extended winds in the early solar system } 

\OWNwebpage{http://www.mso.anu.edu.au/~raquel/} 
\keys{{chondrule} {formation} {solar} {system} {star} {protostellar} {disks}} 

\hypersetup{
    baseurl={\OWNwebpage}, 
    citecolor=black, 
    colorlinks=true, 
    draft=false, 
    filecolor=black, 
    linkcolor=black, 
    pdfauthor={\name}, 
    pdfcreator={\name \ by \ PDFLaTex}, 
    pdfdisplaydoctitle=true, 
    pdffitwindow=true, 
    pdfkeywords={\keys}, 
    pdfsubject={Abstract for LPSC, copyright \@\ \number\the\year}, 
    pdftitle={\DocTitle}, 
    plainpages=false, 
    urlcolor=black 
} 

\begin{document}

\titlearea{\DocTitle}{\name $^{1,2}$, Trevor Ireland$^{2}$, $^{1}$Research School of Astronomy and Astrophysics, The Australian National University; \href{mailto:\ownemail}{\ownemail}; $^{2}$Research School of Earth Sciences, The Australian National University.}
%

\section{Introduction}
The nature of the thermal processing responsible for chondrules and refractory inclusions \cite{PW05} in the early solar nebula is a long-standing puzzle in planetary science.  
The refractory
mineral assemblages, melt textures, and fractionation of magnesium and silicon isotopes imply
that the precursors experienced temperatures up to 1700-2000 K \cite{G10, RDEH02, R00}. 
The meteoritic evidence for a hot
solar nebula provided by chondrules and refractory inclusions is, however, seemingly at odds
with astrophysical observations. These strongly indicate that protostellar disks -- the inspiralling
disks of gas and dust out of which stars and planets form -- are relatively cool \cite{BCWW00}, 
with typical temperatures that are insufficient to melt and vapourise silicate minerals
at the radial distances sampled by chondrule-bearing meteorites in
the main asteroid belt ($\sim$2.1 -- 3.3 AU).

Several scenarios have been proposed for chondrule formation, including melting and condensation of melts in the high-ambient temperature of the solar nebula, lightning strikes, protoplanetary collisions, bow shocks of highly eccentric planetesimals, bipolar jets, and shock waves (e.g.~see the reviews by \cite{jones00a, R00, SK05b} and references therein). Here we show that refractory objects could have been thermally processed in a radially-extended wind, accelerated magnetically from the surfaces of a protostellar disk. Refractory precursor aggregates are heated while being lifted in the wind, grow through amalgamation, and eventually become heavy enough to drop back to the disk, where they assemble with the matrix. Processing at radial distances of about 1-3 AU can produce temperatures in the appropriate regime to melt chondrules and explain their basic properties, while retaining association with the colder material that provides the chondrite matrix. This mechanism is very general, as these energetic winds are commonly associated with stellar formation. 

\section{Protostellar winds}
\begin{figure}[t!]
\begin{center}
\includegraphics[scale=0.25]{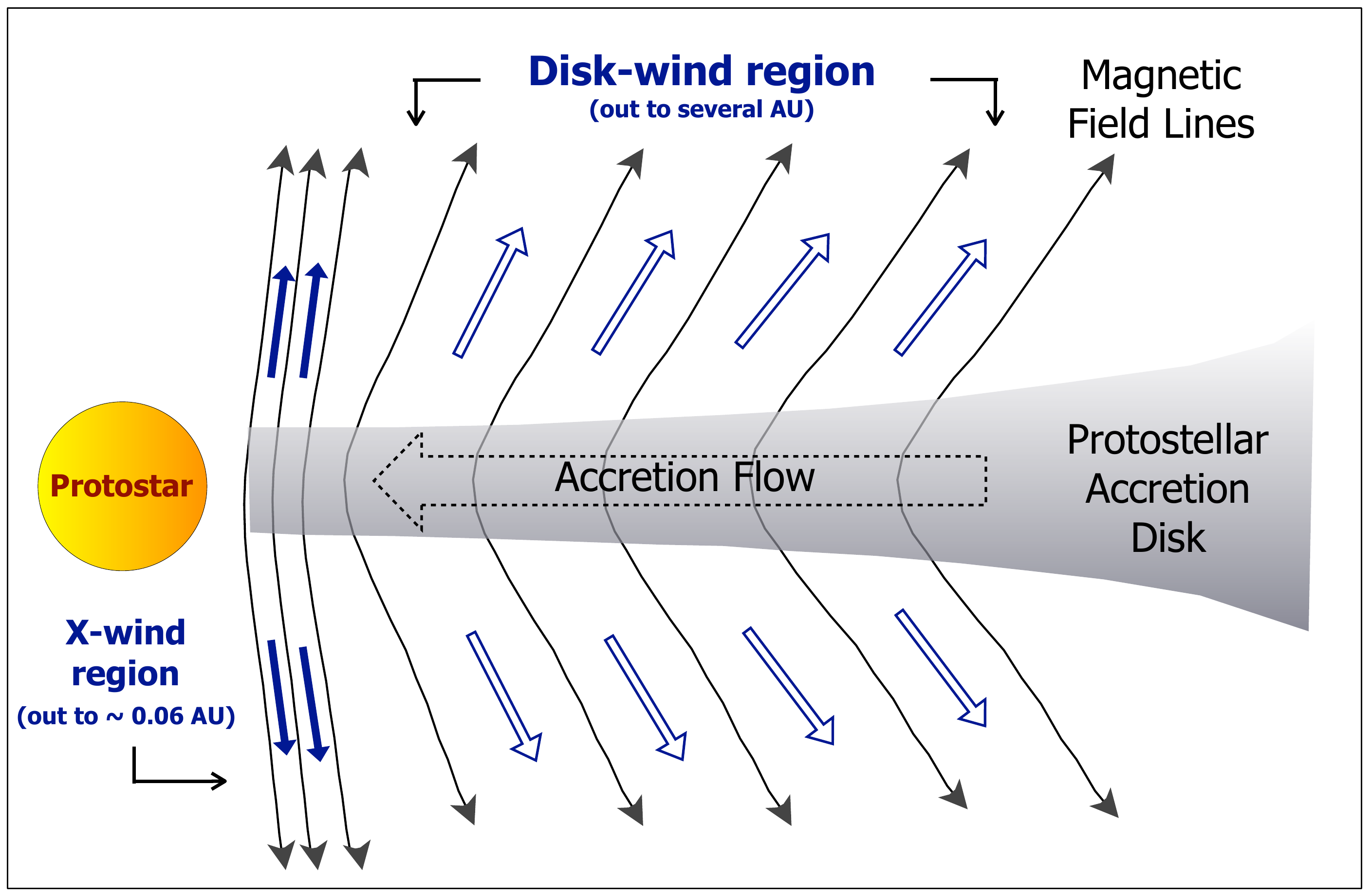}
\vspace{-1.0em}
\caption{\small{Not-to-scale diagram of a protostellar disk, showing the extended Ôdisk-windÕ \protect\cite{BP82} and ÔX-windÕ \protect\cite{shu01a} regions. 
}}
\label{fig:wind}
\vspace{-1.2em}
\end{center}
\end{figure}
Young stars often exhibit powerful outflows that accelerate along the polar axis of the protostellar system \cite{BRD07}.  
These may originate in the disk-star interface (\cite{LB96, shu01a}; `X-winds') or the circumstellar disk itself (`disk winds'; \cite{BP82} and reviews by \cite{POFB07, KS11}).  While X-winds operate very close to the star, near the disk truncation region (within $\sim$ 0.06 AU), disk-winds operate in radially-extended regions around the protostar (see Fig.~\ref{fig:wind}). 
In a X-wind chondrule-formation scenario, chondrule and CAI precursors are heated close to the Sun and thrown outward into the disk.  Issues have been raised with this type of model, with recent work pointing out that many chondrite features seem not to be naturally supported (e.g. \cite{krot09a, DMCB10}).

  An alternative scenario, considered here, is that chondrules are formed within laterally-extended disk winds. %
  These winds 
  are thought to enable the flow of material to the centre to form the star, by removing angular momentum from disk matter. In this scenario, the disk is threaded by large-scale magnetic field
lines, which bend radially outwards. If they are sufficiently inclined, matter near the disk
surface can be accelerated centrifugally along the field lines (the `bead-on-a-wire' effect)
initiating a large-scale outflow. This wind-launching mechanism is supported both by
observational evidence of star-forming regions \cite{GRM06} as well as models of disk formation \cite{KK02}. The correlation between outflow and accretion signatures in these systems \cite{HEG95, Ca07a} strengthens this interpretation. 
  Our models (Salmeron \& Ireland, {\em submitted}) indicate that disk-winds are able to produce the high-temperature processing required for chondrule formation at the distances sampled by chondrite meteorites. 
  
\section{Chondrule processing in disk-winds}
We first obtained a suitable wind solution (see modelling details in \protect\cite{WK93, KSW10, SKW11}) for typical fluid conditions at 1 AU (e.g.~surface density = 900 g cm$^{-2}$, Alfv\'en to sound speed ratio = 0.8, inward speed at the midplane = 0.09 of the sound speed). 
\begin{figure}[t!]
\begin{center}
\includegraphics[scale=0.75]{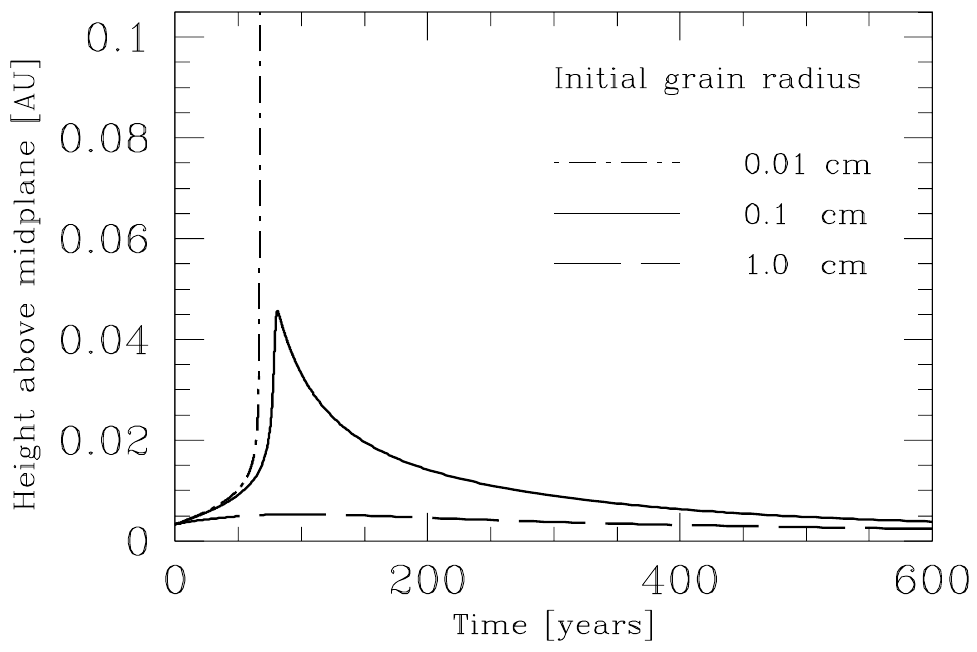}
\vspace{-1.0em}
\caption{\small{Height above the disk mid-plane, as a function of time, for dust particles of the indicated initial sizes, released at their respective terminal velocities at a height of 0.003 AU in a typical disk wind solution.}}
\label{fig:grains}
\vspace{-1.2em}
\end{center}
\end{figure}
We
modelled the motion and growth of spherical, compact  dust particles of different sizes released at
terminal velocity in the wind (Fig.~\ref{fig:grains}). As the particles move through the disk they grow by collisions with smaller grains (for simplicity, we ignore fragmentation). Very small particles (with an initial radius $\sim$ 0.01 cm) are rapidly
accelerated away from the disk by the outflow. More massive, cm-sized particles
quickly sink back to the mid-plane, as the gravitational force dominates over the drag exerted
by the outflowing gas. There is, however, an intermediate size regime (particles with initial
radius $\sim$ 0.1 cm) for which the grain is initially lifted by the wind but, as it grows, the
downward pull of gravity becomes dominant and causes the particle to turn back towards the
disk. 
As it returns to the disk, it may
reassemble with the relatively cold material of the disk interior. 

\begin{figure}[t!]
\begin{center}
\includegraphics[scale=0.55]{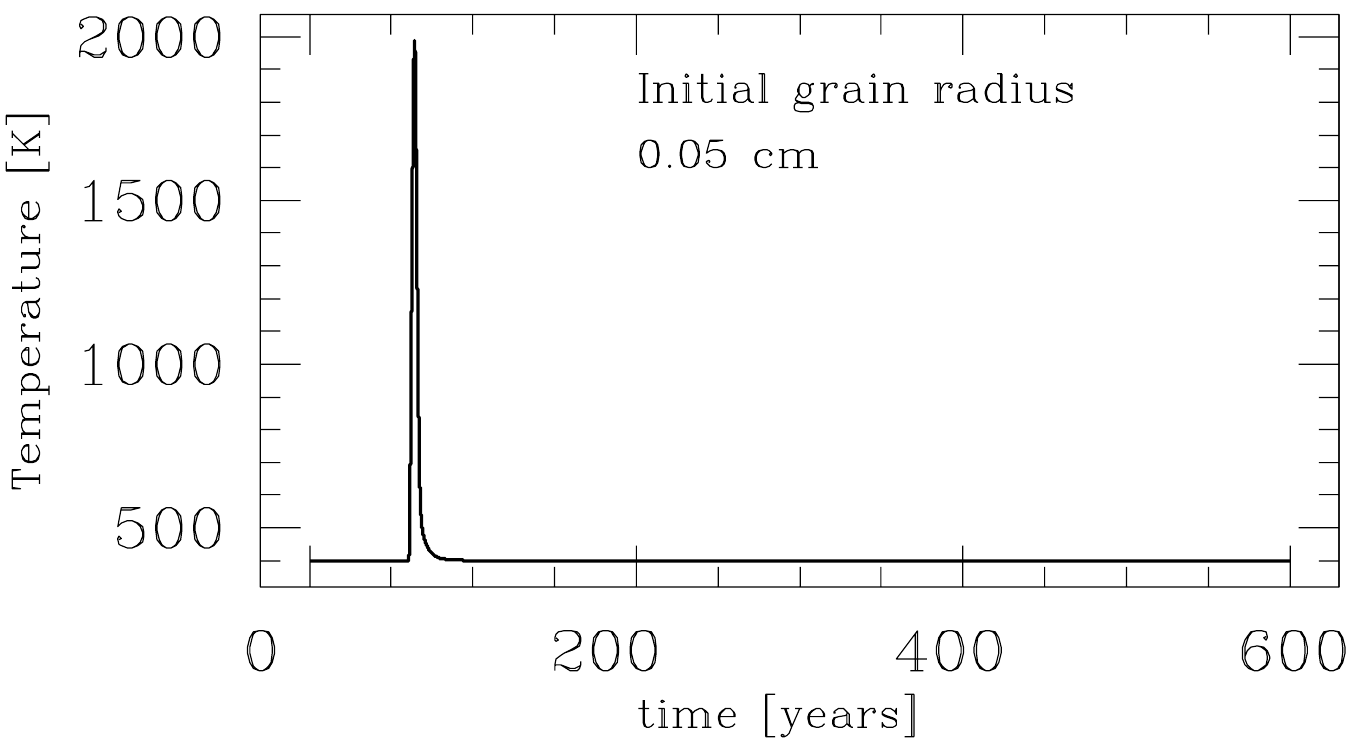}
\vspace{-1.0em}
\caption{\small{Temperature as a function of time, of a dust particle with an initial radius of 0.05 cm, released at 0.003 AU in a typical disk wind solution. The grain initial temperature is 400 K. It experiences a peak temperature near 2000 K and cools over a period of several days.}}
\label{fig:chon}
\vspace{-1.2em}
\end{center}
\end{figure}
In order to illustrate the thermal evolution of a grain in this scenario, we computed the
temperature of a 0.05-cm particle embedded in the wind (Fig.~\ref{fig:chon}). As it moves through the gas, its temperature changes in response to interactions with the gas and the external radiation field \cite{HH91}. We further assumed radiative equilibrium, so that radiative losses are balanced by radiation
absorbed from adjacent particles, as appropriate in dust-rich regions \cite{HH91}. The grainÕs initial temperature is 400 K. 
The gas-dust drift velocity
remains subsonic for most of its trajectory, as both gas and particles accelerate. However, as the
grain grows, it slows down and turns downwards, causing the dust-gas relative velocity and
associated heating to increase dramatically. This particle reaches $\sim$ 0.2 AU above the
disk, where it has grown to 0.08 cm, and experiences a maximum temperature of $\sim$ 2000 K,
both typical values of chondrule material. The grain then cools as it falls back to the disk.

\section{Conclusions}
The outlined disk-wind processing model has attractive features in explaining basic chondrule and chondrite properties. It naturally explains the tight 
size range of chondrules in a given chondrite and their peak temperatures. It can also account for  the large proportion of chondrule material in
chondrites (which follows from continued operation of the wind), the formation of compound chondrules (as the precursors can be
repeatedly `caught' in the wind) and the observed dusty rims of some samples (accreted as the
chondrule returns to the disk interior). This mechanism, being local, may also help explain the long-standing
conundrum presented by the observed complementary composition of chondrules and their
matrix \cite{HP08, P16}. The calculated heating and cooling rates ($\sim$ 1 K/hr) are in qualitative
agreement with inferred rates \cite{C05}. These rates could be
substantially faster if the particle moves into a lower density region so that radiative
equilibrium does not hold. Finally, formation sites at radii of $\sim$ 1-3 AU are consistent with the
observed presence of dust embedded in the gas phase \cite{MCHD03}. 
Disk winds constitute a promising mechanism
by which material could have  been thermally processed in the solar nebula at distances out to the
position of the proto-asteroid belt. More detailed modelling is warranted to further explore it.

\begin{flushleft}\small{\bibliographystyle{LPSCshort_url}
\bibliography{library}}\end{flushleft}

\thispagestyle{empty} 
\end{document}